\documentclass{elsart}
\usepackage{graphicx}
\usepackage{bm}
\usepackage{amssymb}

\begin{document}

\begin{frontmatter}

\title{Spin-flip process through a quantum dot coupled to ferromagnetic electrodes}
\author{Huanwen Lai},
\author[]{Xuean Zhao}, 
\author[]{Zhu-An Xu}, and
\author[]{You-Quan Li}

\address{Zhejiang Institute of Modern Physics and Department of Physics, Zhejiang University,
Hangzhou 310027, China}

\begin{abstract}
We study the spin-dependent transport through a quantum dot
coupled to two ferromagnetic electrodes using the equation of
motion method for the nonequilibrium Green's functions. Our
results show that the conductance and the density of states (DOS)
are strongly dependent on the configurations of the magnetic
electrodes. In parallel configuration of magnetic electrodes the
conductance is affected by the spin-flip process and the Coulomb
repulsion on the dot. The Kondo peak for spin-dependent transport
is splitted into two peaks by the spin-flip process. The locations
of the two peaks are symmetric about no spin-flip peak and the
separation of the splitting is dependent on the strength of the
spin-flip parameter $R$. This effect may be useful to realize the
spin-filter device.
\end{abstract}

\end{frontmatter}

\section{Introduction}
In metallic bulk the dilute magnetic impurities are screened by
the itinerant electron spins and cause anomalous resonant
scattering of conducting electrons\cite{hew}. It shows up a
minimum in electric resistivity at the low temperature. In the
past decade Kondo phenomena in quantum dots coupled to metallic
leads has attracted much
attention\cite{gla,sas,weil,gor,win,bar,barnas} and has been
studied intensively in both theories and experiments. There is a
fundamental difference between the Kondo effect in quantum dots
and the Kondo effect in alloys with magnetic impurities. In the
latter case the dressed localized magnetic moment increases the
scattering cross section and as a result there is a minimum in the
resistivity, whereas for the former case the resistivity depends
on the parity of the level occupied by electrons. However, the
physics responsible for these resistances is similar. In both
cases the effect is due to the spin correlated transport at the
Fermi level.

The Kondo effect has been understood in the structure that a
quantum dot coupled to the normal electrodes in equilibrium and
out of
equilibrium\cite{meir,mei,niu,ser,jau,zhu,giu,ut,kog,zhang}. It is
a consequence of high order perturbation and co-tunnelling process
caused by the electron-spin interaction between electrons in the
quantum dot and in lead. The Kondo effect through quantum dots
coupled to external leads can even induce current in the Coulomb
blockade regime by making spin correlated singlet and triplet with
electron spins outside the quantum dot\cite{sas,weil}. The
conductance is related to the number of electrons in the dot.
Usually the Kondo effect appears when the dot has odd number of
electrons (total spin 1/2), but is absent when the dot has even
number of electrons (total spin 0). The conductance or the density
of states (DOS) of the dot is enhanced by the Kondo effect at low
temperature.  However, for the ferromagnetic-quantum
dot-ferromagnetic system, the question is open. We want to know
what is the manifestations of the Kondo effect when the magnetic
moments of the electrodes are taken into account in finite and
infinite Coulomb repulsion.

In this work, we investigate the spin-dependent transport in an
interacting quantum dot coupled to two ferromagnetic electrodes.
Different from the conventional Kondo problem in normal
electrodes, i.e., quantum dot-normal electrodes system, here we
deal with the strong electronic correlation, which is sensitive to
the relative orientations of magnetization between the two
electrodes. The paper is organized as follows. In Sec.
\ref{sec:model} we describe the model of the system and then
calculate the nonequilibrium Green's functions using the equation
of motion method. The general expression for the electric current
flowing through the dot in terms of nonequilibrium Green's
functions is given. In Sec. \ref{sec:numerical results} we present
numerical results with and without the spin-flip process in the
dot for the arbitrary Coulomb repulsion U. We have some
conclusions in Sec. \ref{sec:conclusion}.

\section{Model}\label{sec:model}
The Hamiltonian for the quantum dot coupled to two ferromagnetic
electrodes reads
\begin{eqnarray}
H &=&\sum_{k;\alpha\in
L,R;\sigma}\varepsilon_{k\alpha\sigma}a^{+}_{k\alpha\sigma}a_{k
\alpha\sigma}
+\sum_{\sigma}\varepsilon_{d}d^{+}_{\sigma}d_{\sigma}
+Ud^{+}_{\uparrow}d_{\uparrow}d^{+}_{\downarrow}d_{\downarrow}
+R(d^{+}_{\uparrow}d_{\downarrow}+d^{+}_{\downarrow}d_{\uparrow})
\nonumber\\ &+& \sum_{k;\alpha\in
L,R;\sigma}(V_{k\sigma}a^{+}_{k\alpha\sigma}d_{\sigma}+ h.c).
\end{eqnarray}
Here the single-particle energy $\varepsilon_{d}$ in quantum dot
is double degenerate in the spin index $\sigma$
($\sigma$=$\uparrow \downarrow, \pm$), and the Coulomb repulsion
interaction in the dot is $U$. The first term in Hamiltonian
describes the free electron in magnetic electrodes, the second and
third terms represent the correlated level of the quantum dot, the
fourth term is to describe spin-orbit coupling which may cause the
spin rotation of an electron in the quantum dot, and the last term
is the spin-dependent hybridization of the quantum dot with the
magnetic electrodes.

Since the spin quantization axes in the electrodes are fixed by
the internal magnetization of the magnets, an electron is in a
superposition of spin-up and spin-down states as it tunnels into
and out the dot. In calculations we have to taken into account the
coherent processes. Technically, we take a
transformation\cite{zhang,cho},
\begin{equation}
d_{\sigma}=\frac{1}{\sqrt{2}}(c_{\sigma}-\sigma
c_{\bar{\sigma}}).\label{e2}
\end{equation}
In terms of Eq. (\ref{e2}) the Hamiltonian becomes
\begin{eqnarray}
H &=& \sum_{k;\alpha\in
L,R;\sigma}\varepsilon_{k\alpha\sigma}a^{+}_{k\alpha\sigma}a_{k
\alpha\sigma}+
\sum_{\sigma}\varepsilon_{c\sigma}c^{+}_{\sigma}c_{\sigma}
+Uc^{+}_{\uparrow}c_{\uparrow}c^{+}_{\downarrow}c_{\downarrow}
\nonumber\\ &+& \sum_{k;\alpha\in
L,R;\sigma}(\frac{1}{\sqrt{2}}V_{k\sigma}a^{+}_{k\alpha\sigma}
c_{\sigma}-\frac{\sigma}{\sqrt{2}}
V^{*}_{k\alpha\sigma}a^{+}_{k\alpha\sigma}c_{\bar{\sigma}}+h.c),
\end{eqnarray}
where $\varepsilon_{c\sigma}$=$\varepsilon_{d}+\sigma R$ for
spin-$\sigma$. The current through the left electrode can be
calculated in terms of nonequilibrium Green's functions with the
tunnelling matrix
\begin{equation}
\bm{V}_{k \alpha}=\frac{1}{\sqrt{2}}\left(
\begin{array}{cc}
V_{k \alpha \uparrow} & V_{k \alpha \downarrow}\\
-V_{k \alpha \uparrow} & V_{k \alpha \downarrow}
\end{array}
\right).
\end{equation}
The current is
\begin{equation}
J=\frac{2e}{\hbar}Re\sum_{k}\int Tr[\bm{V}_{k L} \bm{G}^{<}_{k
L}(\omega)],
\end{equation}
where $[\bm{G}^{<}_{k L}(t)]_{\sigma \sigma'}$=$i\langle a^{+}_{k
L \sigma}(0),c_{\sigma'}(t)\rangle$ is the less Green's function
of the left electrode.

Now we apply Dyson's equation to express the current J by the
Green's functions $\bm{G}$ of the dot,
\begin{equation}
J=\frac{ie}{\hbar}\int\frac{d\epsilon}{2
\pi}Tr\{\bm{\Gamma}^{L}(\omega)
\bm{G}^{<}(\omega)+f_{L}(\epsilon)[\bm{G}^{R}(\omega)-
\bm{G}^{A}(\omega)]\}, \label{current}
\end{equation}
where the less Green's function of the dot
$[\bm{G}^{<}(t)]_{\sigma \sigma'}$=$i\langle c^{+}_{
\sigma'}(t),c_{\sigma}(0)\rangle$. The retarded (advanced) Green's
function of the dot  $[\bm{G}^{R (A)}(t)]_{\sigma \sigma'}$=$\mp$
i$\langle \{ c_{\sigma}(t),c^{+}_{\sigma'}(0)\}\rangle$,
respectively. The upper sign is for retarded Green's function and
the lower sign is for advanced Green's function.
$f_{\alpha}(\epsilon)$ is the Fermi distribution function in the
$\alpha$ electrode, and
\begin{equation}
{\bm{\Gamma}} ^{\alpha }(\omega )=\frac{1}{2}\left(
\begin{array}{cc}
\Gamma _{\uparrow }^{\alpha }(\omega )+\Gamma _{\downarrow
}^{\alpha }(\omega ) & \Gamma _{\downarrow }^{\alpha }(\omega
)-\Gamma
_{\uparrow }^{\alpha }(\omega ) \\
\Gamma _{\downarrow }^{\alpha }(\omega )-\Gamma _{\uparrow
}^{\alpha }(\omega ) & \Gamma _{\uparrow }^{\alpha }(\omega
)+\Gamma _{\downarrow }^{\alpha }(\omega )
\end{array}
\right)
\end{equation}
is the linewidth matrix with $\Gamma^{\alpha}_{\sigma}(\omega)$=$2
\pi \sum_{k} \mid V_{k \alpha \sigma} \mid^{2}
\delta(\omega-\varepsilon_{k \alpha \sigma})$. The spin dependence
of $\Gamma^{\alpha}_{\sigma}(\epsilon)$ originates from the bulk
magnetization of the electrodes.

In order to determine the current J, we should first determine the
Green's function of the dot. As we know the retarded Green's
function $\bm{G}^{R}$ is a conjugation of the advanced Green's
function $\bm{G}^{A}$, we need only to calculate $\bm{G}^{R}$.
With the notation as usual $G^{R}_{\sigma\sigma'}\equiv\ll
c_{\sigma},c^{+}_{\sigma'}\gg\equiv-i\int_{0}^{\infty}e^{i \omega
t}\langle \{ c_{\sigma}(t),c^{+}_{\sigma'}(0)\}\rangle dt$, we
apply the equation of motion method again to calculate the Green's
function $G^{R}_{\sigma\sigma'}$. The first equation of motion for
the retarded Green's function $G^{R}_{\sigma\sigma'}$ is
\begin{equation}
\omega\ll c_{\sigma},c^{+}_{\sigma'}\gg=\langle \{
c_{\sigma},c^{+}_{\sigma'}\}\rangle + \ll
[c_{\sigma},H],c^{+}_{\sigma'}\gg. \label {eq1}
\end{equation}
Using the Eq. (\ref{eq1}), $G^{R}_{\sigma\sigma'}$ can be written
as
\begin{equation}
(\omega-\varepsilon_{c\sigma})G^{R}_{\sigma\sigma'}=
\delta_{\sigma\sigma'}+UG^{(2)}_{\sigma\sigma'}
+\sum_{k;\alpha}\frac{1}{\sqrt{2}}V^{*}_{k\alpha\sigma}
\Gamma^{\sigma\sigma'}_{k\alpha}-
\sum_{k;\alpha}\frac{\bar{\sigma}}{\sqrt{2}}V^{*}_{k\alpha\bar{\sigma}}
\Gamma^{\bar{\sigma}\sigma'}_{k\alpha}, \label{gr01}
\end{equation}
where the second order function $G^{(2)}_{\sigma\sigma'}$ ,
$\Gamma^{\sigma\sigma'}_{k\alpha}$ and
$\Gamma^{\bar{\sigma}\sigma'}_{k\alpha}$ are
$G^{(2)}_{\sigma\sigma'}=\ll
c_{\sigma}c^{+}_{\bar{\sigma}}c_{\bar{\sigma}},c^{+}_{\sigma'}\gg,
\Gamma^{\sigma\sigma'}_{k\alpha}=\ll
a_{k\alpha\sigma},c^{+}_{\sigma'}\gg$, and $
\Gamma^{\bar{\sigma}\sigma'}_{k\alpha}=\ll
a_{k\alpha\bar{\sigma}},c^{+}_{\sigma'}\gg,$ respectively.

There are three new Green's functions on the right hand side of
Eq.(\ref{gr01}). The equation of motion for
$\Gamma^{\sigma\sigma'}_{k\alpha}$ and
$\Gamma^{\bar{\sigma}\sigma'}_{k\alpha}$ are close since the only
new Green's functions are $G^{R}_{\sigma\sigma'}$ and
$G^{R}_{\bar{\sigma}\sigma'}$,
\begin{eqnarray}
\Gamma^{\sigma\sigma'}_{k\alpha}=\frac{1}{\sqrt{2}}V_{k\alpha\sigma}
g^{r}_{1\sigma}G^{R}_{\sigma\sigma'}-
\frac{\sigma}{\sqrt{2}}V_{k\alpha\sigma}g^{r}_{1\sigma}G^{R}_{\bar{\sigma}\sigma'},\\
\Gamma^{\bar{\sigma}\sigma'}_{k\alpha}=\frac{1}{\sqrt{2}}
V_{k\alpha\bar{\sigma}}g^{r}_{2\bar{\sigma}}G^{R}_{\bar{\sigma}\sigma'}-
\frac{\bar{\sigma}}{\sqrt{2}}V_{k\alpha\bar{\sigma}}g^{r}_{2\bar{\sigma}}
G^{R}_{\sigma\sigma'},
\end{eqnarray}
where
$g^{r}_{1\sigma}$=$(\omega-\varepsilon_{k\alpha\sigma}+io^{+})^{-1}$
and
$g^{r}_{2\bar{\sigma}}$=$(\omega-\varepsilon_{k\alpha\bar{\sigma}}+io^{+})^{-1}$.
Note that $g^{r}_{1\sigma}$=$g^{r}_{2\sigma}$.

However, the equation of motion for the new Green's function
$G^{(2)}_{\sigma\sigma'}$, which is generated by the Coulomb
interaction and involves a two-particle Green's function, is not
close. We have to truncate the equations. An approximate solution,
valid for temperatures higher than the Kondo
temperature\cite{lac}, $k_{B}T_{K}\simeq
(U\Gamma)^{(1/2)}$exp$(-\pi|\mu-\epsilon_{\alpha}|/\Gamma)$, is
obtained by neglecting terms in the equation of motion for
$G^{(2)}_{\sigma\sigma'}$ which involves correlations in the
leads\cite{mei}. We can write out $G^{(2)}_{\sigma\sigma'}$ as
follows,
\begin{eqnarray}
(\omega-\varepsilon_{c\sigma}-U)G^{(2)}_{\sigma\sigma'}=n_{\bar{\sigma}}\delta_{\sigma\sigma'}+
\sum_{k;\alpha}\frac{1}{\sqrt{2}}V^{*}_{k\alpha\sigma}\Gamma^{(2)}_{1}-
\sum_{k;\alpha}\frac{\bar{\sigma}}{\sqrt{2}}V^{*}_{k\alpha\bar{\sigma}}\Gamma^{(2)}_{2}\nonumber\\-
\sum_{k;\alpha}\frac{1}{\sqrt{2}}V_{k\alpha\bar{\sigma}}\Gamma^{(2)}_{3}+
\sum_{k;\alpha}\frac{\sigma}{\sqrt{2}}V^{*}_{k\alpha\sigma}\Gamma^{(2)}_{4}+
\sum_{k;\alpha}\frac{1}{\sqrt{2}}V^{*}_{k\alpha\bar{\sigma}}\Gamma^{(2)}_{5}\nonumber\\-
\sum_{k;\alpha}\frac{\sigma}{\sqrt{2}}V^{*}_{k\alpha\sigma}\Gamma^{(2)}_{6},
\end{eqnarray}
with the new notations: $\Gamma^{(2)}_{1}$=$\ll
a_{k\alpha\sigma}c^{+}_{\bar{\sigma}}c_{\bar{\sigma}},c^{+}_{\sigma'}\gg$,
$\Gamma^{(2)}_{2}$=$\ll
a_{k\alpha\bar{\sigma}}c^{+}_{\bar{\sigma}}c_{\bar{\sigma}},c^{+}_{\sigma'}\gg$,
$\Gamma^{(2)}_{3}$=$\ll
c_{\sigma}a^{+}_{k\alpha\bar{\sigma}}c_{\bar{\sigma}},c^{+}_{\sigma'}\gg$,
$\Gamma^{(2)}_{4}$=$\ll
c_{\sigma}a^{+}_{k\alpha\sigma}c_{\bar{\sigma}},c^{+}_{\sigma'}\gg$,
$\Gamma^{(2)}_{5}$=$\ll
c_{\sigma}c^{+}_{\bar{\sigma}}a_{k\alpha\bar{\sigma}},c^{+}_{\sigma'}\gg$,
$\Gamma^{(2)}_{6}$=$\ll
c_{\sigma}c^{+}_{\bar{\sigma}}a_{k\alpha\sigma},c^{+}_{\sigma'}\gg$.
$n_{\bar{\sigma}}$=$\langle c^{+}_{\bar{\sigma}}c_{\bar{\sigma}}
\rangle$ must be calculated self-consistently\cite{niu}.

As for $\Gamma^{(2)}_{i}$, we adopt the truncation as Y. Meir et
al\cite{meir,mei} and write out $\Gamma^{(2)}_{i}$ explicitly as,
\begin{equation}
\Gamma^{(2)}_{1}=\frac{1}{\sqrt{2}}V_{k\alpha\sigma}g^{r}_{1\sigma}G^{(2)}_{\sigma\sigma'}-
\frac{\sigma}{\sqrt{2}}V_{k\alpha\sigma}g^{r}_{1\sigma}f(\varepsilon_{k\alpha\sigma})G^{R}_{\sigma\sigma'},
\end{equation}
\begin{equation}
\Gamma^{(2)}_{2}=-\frac{\bar{\sigma}}{\sqrt{2}}V_{k\alpha\bar{\sigma}}g^{r}_{2
\bar{\sigma}}G^{(2)}_{\sigma\sigma'}+
\frac{1}{\sqrt{2}}V_{k\alpha\bar{\sigma}}g^{r}_{2\bar{\sigma}}
f(\varepsilon_{k\alpha\bar{\sigma}})G^{R}_{\bar{\sigma}\sigma'},
\end{equation}
\begin{eqnarray}
\Gamma^{(2)}_{3}=-\frac{1}{\sqrt{2}}V^{*}_{k\alpha\bar{\sigma}}g^{r}_{3\bar{\sigma}}
G^{(2)}_{\sigma\sigma'}-
\frac{\bar{\sigma}}{\sqrt{2}}V^{*}_{k\alpha\bar{\sigma}}g^{r}_{3\bar{\sigma}}G^{(2)}_{
\bar{\sigma}\sigma'}+
\frac{1}{\sqrt{2}}V^{*}_{k\alpha\bar{\sigma}}g^{r}_{3\bar{\sigma}}f(\varepsilon_{k
\alpha\bar{\sigma}})G^{R}_{\sigma\sigma'}\nonumber\\+
\frac{\bar{\sigma}}{\sqrt{2}}V^{*}_{k\alpha\bar{\sigma}}g^{r}_{3\bar{\sigma}}f(
\varepsilon_{k\alpha\bar{\sigma}})G^{R}_{\bar{\sigma}\sigma'},
\end{eqnarray}
\begin{eqnarray}
\Gamma^{(2)}_{4}=\frac{\sigma}{\sqrt{2}}V^{*}_{k\alpha\sigma}g^{r}_{4\sigma}G^{(2)}_{
\sigma\sigma'}+
\frac{1}{\sqrt{2}}V^{*}_{k\alpha\sigma}g^{r}_{4\sigma}G^{(2)}_{\bar{\sigma}\sigma'}
-\frac{\sigma}{\sqrt{2}}V^{*}_{k\alpha\sigma}g^{r}_{4\sigma}f(\varepsilon_{k\alpha\sigma})
G^{R}_{\sigma\sigma'}\nonumber\\-
\frac{1}{\sqrt{2}}V^{*}_{k\alpha\sigma}g^{r}_{4\sigma}f(\varepsilon_{k\alpha\sigma})G^{R}_{
\bar{\sigma}\sigma'},
\end{eqnarray}
\begin{equation}
\Gamma^{(2)}_{5}=\frac{1}{\sqrt{2}}V_{k\alpha\bar{\sigma}}g^{r}_{5\bar{\sigma}}G^{(2)}_{
\sigma\sigma'}-
\frac{1}{\sqrt{2}}V_{k\alpha\bar{\sigma}}g^{r}_{5\bar{\sigma}}f(\varepsilon_{k\alpha\bar{
\sigma}})G^{R}_{\sigma\sigma'},
\end{equation}
\begin{equation}
\Gamma^{(2)}_{6}=-\frac{\sigma}{\sqrt{2}}V_{k\alpha\sigma}g^{r}_{6\sigma}G^{(2)}_{
\sigma\sigma'}+
\frac{\sigma}{\sqrt{2}}V_{k\alpha\sigma}g^{r}_{6\sigma}f(\varepsilon_{k\alpha\sigma})
G^{R}_{\sigma\sigma'},
\end{equation}
with the notations as
$g^{r}_{3\bar{\sigma}}$=$(\omega-\varepsilon_{c\sigma}-\varepsilon_{c\bar{\sigma}}+
\varepsilon_{k\alpha\bar{\sigma}}-U+io^{+})^{-1}$,
$g^{r}_{4\sigma}$=$(\omega-\varepsilon_{c\sigma}-\varepsilon_{c\bar{\sigma}}+
\varepsilon_{k\alpha\sigma}-U+io^{+})^{-1}$,
$g^{r}_{5\bar{\sigma}}$=$(\omega-\varepsilon_{c\sigma}+\varepsilon_{c\bar{\sigma}}-
\varepsilon_{k\alpha\bar{\sigma}}+io^{+})^{-1}$, and
$g^{r}_{6\sigma}$=$(\omega-\varepsilon_{c\sigma}+\varepsilon_{c\bar{\sigma}}-
\varepsilon_{k\alpha\sigma}+io^{+})^{-1}$. Notice that
$g^{r}_{3\sigma}$=$g^{r}_{4\sigma}$.

Defining $\Sigma_{i\sigma}$=$\sum_{k;\alpha}\frac{1}{2}\mid
V_{k\alpha\sigma}\mid^{2}g^{r}_{i\sigma}$ and
$\Sigma^{f}_{i\sigma}$=$\sum_{k;\alpha}\frac{1}{2}\mid
V_{k\alpha\sigma}\mid^{2}f(\varepsilon_{k\alpha\sigma})g^{r}_{i\sigma}$,
we have $G^{(2)}_{\sigma\sigma'}$ in the form of
\begin{eqnarray}
(\omega-\varepsilon_{c\sigma}-U)G^{(2)}_{\sigma\sigma'}=
n_{\bar{\sigma}}
\delta_{\sigma\sigma'}+(\Sigma_{1\sigma}+\Sigma_{2\bar{\sigma}}+\Sigma_{3
\bar{\sigma}}+\Sigma_{4\sigma}
+\Sigma_{5\bar{\sigma}}\nonumber\\+\Sigma_{6\sigma})G^{(2)}_{\sigma\sigma'}+
(\bar{\sigma}\Sigma_{3\bar{\sigma}}+\sigma\Sigma_{4\sigma})G^{(2)}_{\bar{\sigma}
\sigma'}-
(\Sigma^{f}_{3\bar{\sigma}}+\Sigma^{f}_{4\sigma}+\Sigma^{f}_{5\bar{\sigma}}\nonumber\\+
\Sigma^{f}_{6\sigma})G^{R}_{\sigma\sigma'}-
(\sigma\Sigma^{f}_{1\sigma}+\bar{\sigma}\Sigma^{f}_{2\bar{\sigma}}+\bar{\sigma}
\Sigma^{f}_{3\bar{\sigma}}+\sigma\Sigma^{f}_{4\sigma})
G^{R}_{\bar{\sigma}\sigma'}.
\end{eqnarray}
Making the truncation as above, the equation for $G^{(2)}_{\sigma
\sigma'}$ is close. It now only involves the Green's functions
$G^{(2)}_{\sigma \sigma'}$, $G^{(2)}_{\bar{\sigma} \sigma'}$,
$G^{R}_{\sigma\sigma'}$ and $G^{R}_{\bar{\sigma}\sigma'}$. One can
solve them self-consistently. With the notations of
$\hat{\varepsilon}_{c\sigma}$=$\left(
\begin{array}{cc}
\varepsilon_{c\sigma}& 0 \\
0 & \varepsilon_{c\bar{\sigma}}
\end{array}
\right)$,  $\bf {\hat{n}}$=$\left(
\begin{array}{cc}
n_{\bar{\sigma}} & 0 \\
0 & n_{\sigma}
\end{array}
\right)$,

$\hat{\Sigma}_{0}$=$\left(
\begin{array}{cc}
\Sigma_{1\uparrow}+\Sigma_{2\downarrow} & \Sigma_{1\downarrow}-\Sigma_{2\uparrow} \\
\Sigma_{1\downarrow}-\Sigma_{2\uparrow} &
\Sigma_{1\uparrow}+\Sigma_{2\downarrow}
\end{array}
\right)$,  $\hat{\Sigma}_{1}$=$\left(
\begin{array}{cc}
\Sigma^{f}_{5\downarrow}+\Sigma^{f}_{6\uparrow} &
\Sigma^{f}_{1\uparrow}-\Sigma^{f}_{2\downarrow} \\
\Sigma^{f}_{1\uparrow}-\Sigma^{f}_{2\downarrow} &
\Sigma^{f}_{5\uparrow}+\Sigma^{f}_{6\downarrow}
\end{array}
\right)$,

$\hat{\Sigma}_{2}$=$\left(
\begin{array}{cc}
\Sigma_{1\uparrow}+\Sigma_{2\downarrow}+\Sigma_{5\downarrow}+\Sigma_{6\uparrow}
& 0 \\ 0 &
\Sigma_{1\downarrow}+\Sigma_{2\uparrow}+\Sigma_{5\uparrow}+\Sigma_{6\downarrow}
\end{array}
\right)$, and

$\hat{\Sigma}^{(f)}_{3}$=$\left(
\begin{array}{cc}
\Sigma^{(f)}_{3\uparrow}+\Sigma^{(f)}_{4\downarrow} &
\Sigma^{(f)}_{3\uparrow}-\Sigma^{(f)}_{4\downarrow} \\
\Sigma^{(f)}_{3\uparrow}-\Sigma^{(f)}_{4\downarrow} &
\Sigma^{(f)}_{3\uparrow}+\Sigma^{(f)}_{4\downarrow}
\end{array}
\right)$, defining $\hat{\Sigma}_{4}$=$\omega
\hat{\bm{1}}-\hat{\varepsilon}_{c\sigma}-U
\hat{\bm{1}}-\hat{\Sigma}_{2}-\hat{\Sigma}_{3}$, one can write the
Green's function $G^{R}$ in a compact form as below
\begin{equation}
\hat{G}^{R}=[\omega \hat{\bm
1}-\hat{\varepsilon}_{c\sigma}-\hat{\Sigma}_{0}+U
\hat{\Sigma}_{4}^{-1}(\hat{\Sigma}^{f}_{3}+\hat{\Sigma}_{1})]^{-1}
(\hat{\bm 1}+U {\bf \hat{n}} \hat{\Sigma}_{4}^{-1}).
\end{equation}
In the infinite $U$ limit, we recover the result in Refs.
\cite{meir,zhang},
\begin{equation}
\hat{G}^{R}=(\omega \hat{\bm
1}-\hat{\varepsilon}_{c\sigma}-\hat{\Sigma}_{0}-\hat{\Sigma}_{1})^{-1}
(\hat{\bm 1}-\bf {\hat{n}}).
\end{equation}

To determine the lesser Green's function of the dot $\bm{G}^{<}$,
we use the Ng's ansatz\cite{ng} in our case. The interaction
lesser and greater self-energies are assumed to be of the form
$\Sigma^{<,>}$=$\Sigma_{0}^{<,>}\bm{A}$, where $\bm{A}$ is a
matrix to be determined by the condition of
$\Sigma^{<}$-$\Sigma^{>}$=$\Sigma^{R}$-$\Sigma^{A}$. This ansatz
is exact in the noninteracting limit($U$=0) and guarantees
automatically the current conservation law. As a result one
obtains
$\Sigma^{<}$=$\Sigma_{0}^{<}[\Sigma_{0}^{R}-\Sigma_{0}^{A}]^{-1}[\Sigma^{R}-\Sigma^{A}]$.
Using this results, $\bm{G}^{<}$ can be obtained by Keldysh
equation $\bm{G}^{<}$=$\bm{G}^{R} \Sigma^{<} \bm{G}^{A}$.
Substituting the expressions of the Green's functions of quantum
dot into Eq. (\ref{current}) and defining
$\bar{\Sigma}^{<}$=$\bm{\Gamma}^{R}(\bm{\Gamma}^{L}+
\bm{\Gamma}^{R})^{-1}(\Sigma^{R}-\Sigma^{A})$, one obtains an
expression of the tunnelling current
\begin{equation}
J=\frac{ie}{\hbar}\int \frac{d\epsilon}{2 \pi} Tr [\bm{\Gamma}^{L}
\bm{G}^{R} \bar{\Sigma}^{<}
\bm{G}^{A}][f_{L}(\epsilon)-f_{R}(\epsilon)].
\end{equation}
This expression generalizes the well know current formula to the
spin-dependent Anderson model with additional spin-flip relaxation
and allows one to describe the coherent spin transport through an
interacting quantum dot coupled to magnetic electrodes.

\section{Numerical results}\label{sec:numerical results}
In this section, we will present the numerical results of
spin-dependent DOS and conductance. For simplicity, we neglect the
energy dependence of the tunnelling matrix elements and consider
parallel and antiparallel configurations. If the magnetic moments
in the electrodes are parallel, the spin-majority electrons are
assumed to be spin-up and the spin-minority electrons are
spin-down. In the antiparallel configuration, the magnetization of
the left electrode is up and the magnetization of the right
electrode is down. With these assumptions the coherent spin
transport parameters can be conveniently expressed  by introducing
magnetic polarization factors $p_{L}$ and $p_{R}$ for the left and
right barriers, respectively. Therefore,
$\Gamma^{L}_{\uparrow(\downarrow)}$=$\Gamma_{0}(1\pm p_{L})$ and
$\Gamma^{R}_{\uparrow(\downarrow)}$=$\alpha\Gamma_{0}(1\pm p_{R})$
are for the parallel configuration, and
$\Gamma^{L}_{\uparrow(\downarrow)}$=$\Gamma_{0}(1\pm p_{L})$,
$\Gamma^{R}_{\uparrow(\downarrow)}$=$\alpha\Gamma_{0}(1\mp p_{R})$
for the antiparallel configuration. $\Gamma_{0}$ describes the
coupling between the quantum dot and the left electrode without
internal magnetization. We express all parameters in the unit of
$\Gamma_{0}$ in the following calculation. $\alpha$ denotes tunnel
asymmetric factor of the left and right barriers. In this work, we
assume the symmetric barriers, i.e., $\alpha$=1,
$p_{L}$=$p_{R}$=$p$.

\begin{figure}[tbph]
\vbox to2.0in{\rule{0pt}{2.0in}} \includegraphics{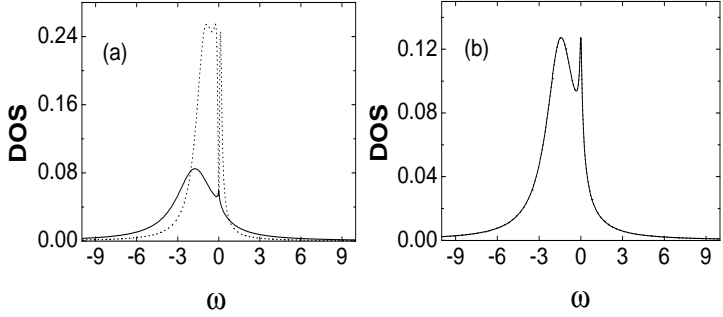}
\caption{Density of states in the parallel (a) and antiparallel
(b) configurations in the infinite $U$ limit, the solid line is
for spin-up and the dotted line is for spin-down. Here
$\varepsilon_{d}$=-2, $K_{B}$T=0.001, p=0.5, R=0.} \label{fig1}
\end{figure}

The spin-resolved DOS in the dot are calculated via the relation
$\rho_{\uparrow (\downarrow)}$ = $-1/\pi Im\{G^{R}_{\uparrow
\uparrow} \mp G^{R}_{\uparrow \downarrow} \mp G^{R}_{\downarrow
\uparrow} +G^{R}_{\downarrow \downarrow}\}$. The Kondo resonance
for each spin is clearly manifested by the peak at $\omega=$0 in
the DOS for both parallel and antiparallel configurations.
However, the shape of the peak is sensitive to the magnetic
configurations of the electrodes.  As observed from Fig.
\ref{fig1}(a), in the parallel configuration the spin-up DOS
(solid line in Fig. \ref{fig1}(a)) is dramatically different from
that of spin-down (dotted line in Fig. \ref{fig1}(a)) in the
infinite $U$ limit, while in the antiparallel configuration the
DOS for spin-up and spin-down are identical, see Fig.
\ref{fig1}(b).  In Fig. \ref{fig1}(b) the Kondo peak for spin-up
and spin-down is not affected by the magnitude of magnetic
polarizations $p$. As we know that the spin current of electrons
through the antiparallel configuration cannot be polarized,
whereas can be polarized through the parallel
configuration\cite{dob}. Therefore, the DOS for spin-up is
different from that of spin-down in the parallel configuration and
is the same as that of spin-down in the antiparallel
configuration.

\begin{figure}[tbph]
\vbox to2.5in{\rule{0pt}{2in}} \includegraphics{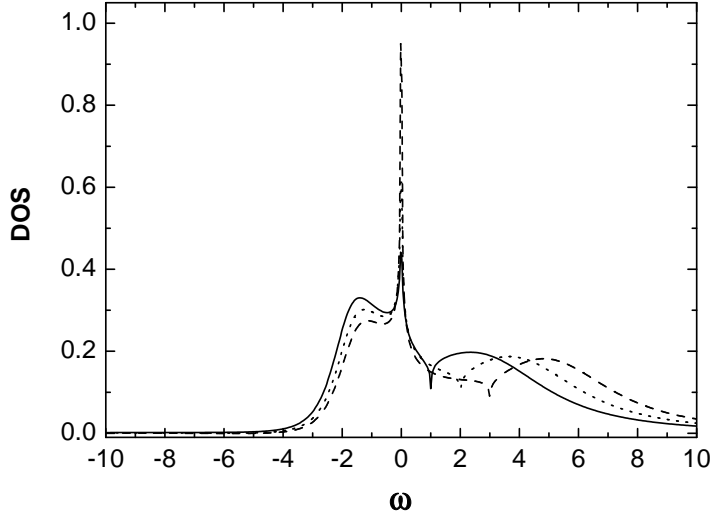}
\caption{Spin-up density of states  in the antiparallel
configuration for $U$=5(solid line), 6(dotted line), and 7(dashed
line). Here R=0, the other parameters are the same  as in Fig.1.}
\label{fig2}
\end{figure}

\begin{figure}[tbph]
\vbox to2.5in{\rule{0pt}{2in}} \includegraphics{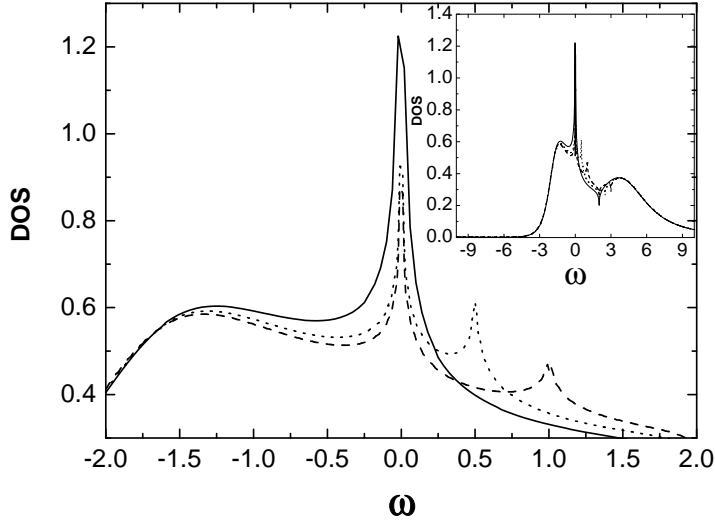}
\caption{Total density of states in the antiparallel configuration
with v=0 (solid line), 0.5 (dotted line), and 1 (dashed line),
respectively. Here $U$=6, the other parameters are the same as in
Fig.1.} \label{fig3}
\end{figure}

In the limited Coulomb repulsion $U$, the spin-up DOS in the
antiparallel configuration are plotted in Fig. \ref{fig2}. It can
be seen from Fig. \ref{fig2}, there are two broad maxima in DOS,
which are centered at the positions of the dot level
$\varepsilon_{d}$  and  $\varepsilon_{d}+U$, respectively.
However, for infinite $U$ limit case,  as shown in Fig.
\ref{fig1}, there is only one broad maximum. It is consistent with
physical intuition since for the infinite $U $ there is only one
level in the dot but are two levels for the finite $U$.  This
result agrees with other authors' work\cite{zhang,ser}. The Kondo
peak in DOS is exactly located at the Fermi level, see the sharp
peak  in Fig. \ref{fig2}. This is the equilibrium situation (the
applied bias is set to be zero). When the voltage is applied (the
left chemical potential is different from that of the right), the
Kondo peak splits into two peaks with lower intensities at
positions of the left and the right chemical potentials. This
behavior is shown in Fig. \ref{fig3} for three different values of
the bias voltages. As the bias voltage increases the splitting of
the Kondo peak becomes wider and is equal to the difference of the
bias. The Fig. \ref{fig3} is plotted in a narrow energy range
around the Kondo peak, the inset of the figure gives a wider view
in energy. It is evident that the peaks occur at the Fermi levels
of the two electrodes. The solid line is for zero bias, there is
only one peak at $\omega=0$. The dotted line is for bias v=0.5 and
the second peak appears at the position $\omega=0.5$. The dashed
line is for v=1, the splitting peak is at $\omega=1$. This
behavior is in accordance with other results\cite{mar,kra,cos}.

\begin{figure}[tbph]
\vbox to2.5in{\rule{0pt}{2in}} \includegraphics{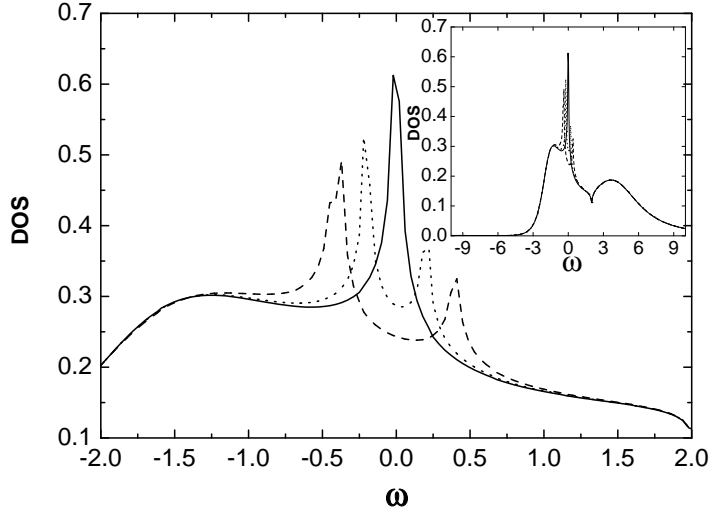}
\caption{Spin-up density of states in the antiparallel
configuration with R=0 (solid line), 0.2 (dotted line), and 0.5
(dashed line), respectively. Here $U$=6, the other parameters are
the same as in Fig.1.} \label{fig4}
\end{figure}

The interesting feature in this work is that there exists a
splitting due to the spin-flip transition processes. In the
preceding discussions there is always a Kondo peak at the energy
$\omega=0$. However, if we
set spin-flip parameter R not to be zero, as shown in Fig.
\ref{fig4} for the DOS of spin-up in the antiparallel
configuration, the original Kondo peak  splits into two peaks. The
two  peaks appear at the positions of $\omega=\pm R$,
respectively, and the original peak at $\omega=0$ disappears. In
Fig. \ref{fig4} the solid line is plotted without spin-flip
transition, the dotted line is for spin-flip transition parameter
$R=0.2$ and the dashed line is $R=0.5$. The inset in Fig.
\ref{fig4} is also a global view of the Kondo effect. The
splitting is due to the coupling of spin-flip transition processes
between the dot and the magnetic electrodes and can be expressed
by the high order self-energy.

\begin{figure}[tbph]
\vbox to2.8in{\rule{0pt}{2in}} \includegraphics{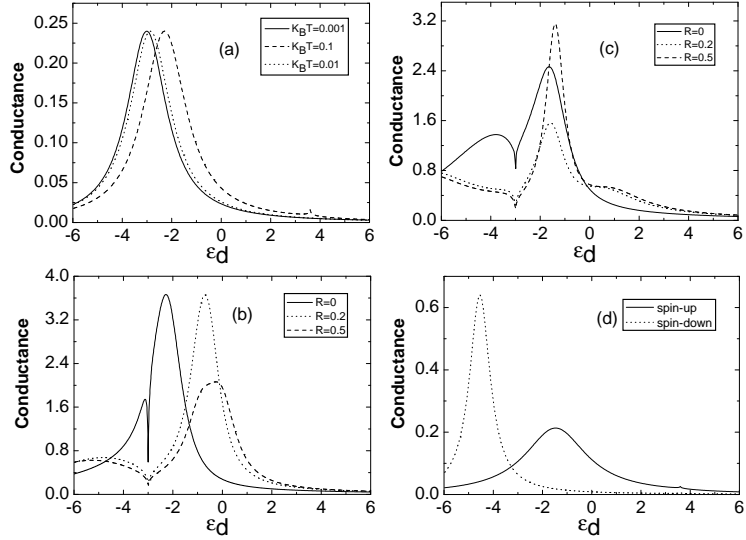} \caption{The
conductance in the antiparallel and antiparallel configurations.
(a) in the antiparallel configuration for the infinite $U$ limit
and different temperatures. The spin-flip parameter R=0. (b) in
the antiparallel configuration for different R, here U=6 and
$K_{B}$T=0.001.  (c) in the parallel configuration for different R
,  the other parameters are the same as in (b).  (d) the
conductance in the parallel configuration for the spin-up and
spin-down, respectively, $K_{B}$T=0.001, the other parameters are
the same as in (a).} \label{fig5}
\end{figure}

In Fig. \ref{fig5}, we show the linear response conductance as a
function of energy $\varepsilon_{d}$ of the dot in the parallel
and antiparallel configurations. In the antiparallel configuration
the conductance is similar to the normal case, i.e., the Kondo
effect broadens the conductance peak, see Fig. \ref{fig5}(a). The
peaks are similar in shape for different temperatures. They only
shift a little to each other. As temperature increases the peak
shifts to the higher energy. This is caused by the shift of the
Fermi energy for different temperatures. Due to the DOS for
spin-up and spin-down are the same in the antiparallel
configuration, as seen in Fig. \ref{fig1}, the conductances for
spin-up and spin-down are overlapped in this configuration.
However, when $U$ is finite, for instance $U$=6, the conductance
of spin-up differs to that of spin-down, see solid line in Fig.
\ref{fig5}(b). If the spin-flip process on the quantum dot is
involved, this shift becomes more dramatically, see the dotted
line and dashed line in Fig. \ref{fig5}(b). For a large value of R
the amplitude of conductance is small. In the parallel
configuration the conductances for spin-up and spin-down are
different\cite{dob}, see Fig. \ref{fig5}(d), the resonance of
spin-up conductance (solid line) has an apparent shift to the
spin-down (dotted line) due to different densities of states for
them [see Fig. \ref{fig1}(a)]. The existence of the spin-flip
transition R reduces the spin-down conductance under the finite
$U$ condition, see Fig. \ref{fig5}(c). Increasing the value of R
the spin-up conductance becomes higher and the spin-down
conductance becomes lower. This effect is caused by the
correlation between the dot and the ferromagnetic electrodes. For
a large enough R the peak of conductance of spin-down is
suppressed  and  the spin-up peak is enhanced. We expect this
result may be important in the exploiting the spin-filter devices.

\section{Conclusion}\label{sec:conclusion}
Using the Anderson model, we study the Kondo effect in the
ferromagnetic-quantum dot-ferromagnetic system. We find that the
spin-dependent transport in such a system is sensitive to the
alignment of the magnetic moments in electrodes. Due to the
external applied bias or the spin-flip process on the dot, the
original Kondo peak splits into two peaks. The conductances
in the parallel and antiparallel magnetic configurations are
affected by the spin-flip transition and the Coulomb repulsion on
the dot. We expect that the effects can be useful to realize
the high spin polarization
devices.

\section*{Acknowledgments}
This work was supported by NSFC Grant No.
10274069, 60471052, 10225419,10347003; the Zhejiang Provincial Natural
Foundation M603193.

\end{document}